\documentclass[doublecol]{epl2}
\usepackage{amssymb,latexsym,mathrsfs}
\usepackage{amsfonts}
\usepackage{amsmath}
\usepackage{graphicx}

\newcommand{\beq}{\begin{equation}}
\newcommand{\eeq}{\end{equation}}
\newcommand{\beqn}{\begin{eqnarray}}
\newcommand{\eeqn}{\end{eqnarray}}
\newcommand{\bearr}{\begin{array}}
\newcommand{\enarr}{\end{array}}

\def\bea{\begin{eqnarray}}
\def\eea{\end{eqnarray}}
\def\ba{\begin{array}}
\def\ea{\end{array}}
\def\n{\nonumber}

\def \prl { Phys. Rev. Lett.}

\title{Restricted Exclusion Processes without Particle Conservation Flows to Directed Percolation}
\author{ Urna Basu and P. K. Mohanty}
\institute{Theoretical Condensed Matter Physics Division, Saha Institute of Nuclear Physics,
1/AF Bidhan Nagar, Kolkata, 700064 India.}
\date{\today}

 \abstract{
Absorbing phase transition in restricted exclusion processes  are  characterized  by 
simple integer exponents. We show that this critical behaviour flows to the directed 
percolation (DP) universality  class when particle conservation  is broken suitably.
The same transition, when  studied using   the average density as the  controlling 
parameter, yields   critical exponents  quite different from DP; we argue   that these 
exponents are actually   related to DP by a scaling factor $1/\beta_{DP}.$  
These conclusions   also apply to  conserved lattice gas in  one dimension.  
}

\begin{document}

\maketitle

Directed percolation(DP)\cite{dp} is  the most  dominant universality  class
of  absorbing phase transition (APT). The famous `DP conjecture' \cite{gras}  claims that 
that any APT  with a fluctuating scalar order parameter  should generically belong 
to  DP class.  
Several models systems,  like reaction diffusion processes\cite{react-diff}, 
depinning transitions\cite{depinning}, damage spreading\cite{spread}, 
synchronization transition\cite{synchro}, sand-pile models\cite{pk}  
and certain probabilistic cellular automata\cite{ca} 
are  known to undergo  APT belonging to  this universality class. Recently  the DP  critical behaviour 
has  been verified  experimentally  in context of liquid crystals\cite{DPexp}.

Surprisingly, in spite  of its robustness  directed percolation  
does not have an exact solution, even in  $(1+1)$-dimension, although 
the critical  exponents are  calculated to  high accuracy\cite{DPjensen}.  
Only a few  models  of APT are solved exactly;  examples  include compact directed 
percolation (CDP)\cite{cdp},  branching-annihilating random walk (BARW)\cite{BARW},
conserved lattice gas (CLG) model in one dimension \cite{Oliveira} and  
restricted asymmetric simple exclusion process (RASEP)\cite{rasep}. 
The non-DP behaviours  in CDP and BARW  owe to  the special symmetries  of the 
order parameter, namely  particle-hole symmetry and parity respectively. 
The scalar order parameters of CLG and  RASEP  do not have any special symmetry 
but it is coupled to  an additional conserved field present in the system. Conventionally, 
presence of this  additional field is  blamed for these models to be in a non-DP class.
This reasoning is consistent with  the non-DP behaviour  observed in  
fixed energy sand pile models\cite{cons-h} and a 
conjecture by  Rossi et. al. \cite{clg} that  critical behaviour of any APT  
in presence of a conserved field  should differ from  DP.

\begin{figure}[t]
 \centering
 \includegraphics[width=8 cm]{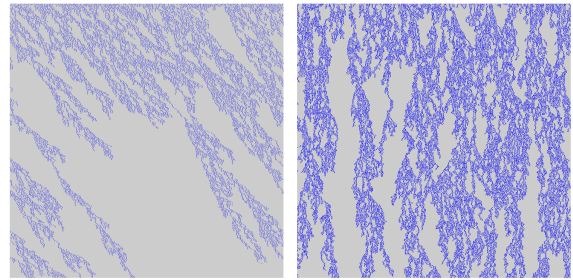}\vspace*{-0.3 cm}
\caption{Propagation of activity in asymmetric (left) and  symmetric (right)    restricted 
exclusion process starting from  the maximally active configuration $\{110110\dots\},$  
for $w=0.560.$ The blue (dark)  pixels  represent active sites. }
 \label{fig:clus}\vspace*{-0.3 cm}
\end{figure}

However, more recent  investigations reveal \cite{nomanna} that  the  actual critical behaviour of  one dimensional fixed energy  sand-pile models belong to DP. The apparent non-DP 
behaviour was only a transient phenomenon; coupling of the order parameter  to the  
additional conserved  height field does not  necessarily induce a new critical 
behaviour. APTs in energy exchange   models\cite{cclf} too 
show DP behaviour even though   the total energy  of the system is
conserved. 
In view of these examples it is not convincing to  accept
`conservation'  as the cause  of an absorbing transition 
not being in DP, unless one explicitly  shows that the 
specific model flows to  DP if the existing conservation 
is broken.

In this article we  study  the restricted asymmetric simple exclusion process (RASEP) 
as an example of a simple  and exactly solvable  non-DP  critical behaviour 
in presence of a conserved density field. The dynamics of 
RASEP is  extended  to incorporate deposition and evaporation 
of particles in the bulk so that the  conservation of density  is  violated;
the system  now  undergoes an APT at a  critical deposition rate $w_c$. 
The resulting  critical behaviour is found to be  in the  DP 
universality class. However, the  critical exponents 
turn out to be much different from DP when the average density 
$\rho$  is   treated as the control parameter, like  the conserved RASEP.
 We show that these exponents are  actually related  to DP by a scaling factor $1/\beta_{DP}$ as  
the  singular   behaviour  of  $\rho$   at the critical point is also 
characterized by  DP  exponents.

\section{Model}

The  restricted asymmetric exclusion process is defined  on a one dimensional periodic 
lattice labeled by sites $i=1,2\dots L.$   The sites are either vacant or occupied by 
at most one particle; corresponding site variables are $s_i=0,1.$  
A particle from a randomly chosen site hops forward to its neighbouring 
vacant site only when it is preceded by at least $\mu$ particles. Thus, the  model with a specific $\mu$ 
evolves according to  the  corresponding particle conserving dynamics,   
\bea
\mu=1 &:& 110 \rightarrow 101\cr
\mu=2 &:& 1110 \rightarrow 1101\cr
\mu=3 &:& 11110 \rightarrow 11101  ~~\dots~~etc.
\label{eq:mu}
\eea

Any configuration which does not have  
more than $\mu$ consecutive particles anywhere in the system is an {\it absorbing} configuration.
If the system reaches such a configuration it gets stuck there forever.
All the other configurations, which have at least one block of more than $\mu$ 
consecutive particles, are active. Since all the configurations  
with particle density $\rho> \mu/(1+\mu)$ are active, the APT is expected to occur at some lower density.
It has been shown  analytically \cite{rasep} that  the system undergoes an APT  at 
$\rho_c =  \mu/(1+\mu)$ exactly with critical exponents 
\bea
\beta=1,\nu_\perp=1,\eta=1,\beta^\prime=0. \label{eq:rasepexp}
\eea
Moreover, numerical simulations  have suggested \cite{Lee} that $\alpha=1/2, z=1$, and $\nu_\parallel =2 $.  

To claim that the conserved density is  the sole cause of
the non-DP behaviour   of RASEP (Eq. (\ref{eq:rasepexp})) we intend to  break   
the conservation of density and  investigate  whether  this  
truly {\em drives} the transition to the DP fixed point. 
However, any arbitrary dynamics that deposits (evaporates) particle at (from) 
a vacant (occupied) site may altogether wipe out the possibility of a phase transition of the system. 
For example,  the simplest way one can break the density conservation is by augmenting 
the dynamics with 
$$
0 \mathop{\leftrightarrows}_{w}^{1-w} 1.
$$
This bulk  dynamics fixes the average density of the system $\rho= w.$  However,  
there are  no absorbing configurations  left anymore as the  dynamics  allows 
any arbitrary configuration  to become active eventually . Thus, the system is  active 
for any $w$  and  does not show  any transition. To preserve the
APT, the non-conserving dynamics must be  {\it designed  suitably}, so that 
the absorbing states are still present in the phase space.  One possible way to ensure  
this is to device  a  dynamics  which does not affect the existing 
absorbing configurations. In other words,  one ought to  deposit or evaporate  particles 
only on  active configurations -  
all the configurations which were absorbing in the conserved model would  then still 
be absorbing in the non-conserved version.

A  possible non-conserving local dynamics for RASEP, which fulfills this condition
for any $\mu,$ is 
\bea
\mu =1 &:& 110 \mathop{\leftrightarrows}_{w}^{1-w} 111, \n \\
\mu =2 &:& 1110 \mathop{\leftrightarrows}_{w}^{1-w} 1111, \n \\
\mu =3 &:& 11110 \mathop{\leftrightarrows}_{w}^{1-w} 11111 \quad \dots \; etc,
\label{eq:110_111}
\eea
where $0\le w\le 1$  fixes  the  density by controlling both the deposition 
and evaporation of particles. Clearly, large $w$  would  correspond to high density. 
This dynamics, when  added to the conserving  evolution rule \eqref{eq:mu}, 
allows addition or removal of a particle  in the neighbourhood of an active site {\it only}. Absorbing configurations, which  do not have  $(\mu+1)$ or more consecutive particles, 
are not  affected by  the non-conserving dynamics  \eqref{eq:110_111}.  

\begin{figure}[t]
 \centering
 \includegraphics[width=6.5 cm]{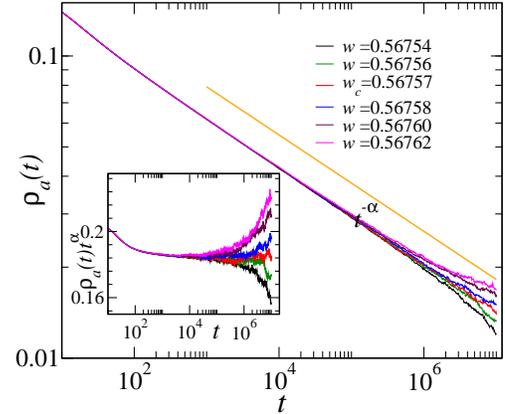} \vspace*{-0.3 cm}
 \caption{Decay of $\rho_a(t)$  from a maximally active configuration 
$\{110110110 \dots \}$  for different values of $w.$  
The same data  is plotted as $\rho_a(t)t^\alpha$ versus  $t$ in the inset. In  the  long time,  this 
curve becomes  horizontal for $w_c=0.56757$ . Here, $L = 2^{16}.$} \vspace*{-0.3 cm}
 \label{fig:wc}
\end{figure}

Although the  number of particles  in the system  fluctuates,  the average density $\rho = \langle s_i \rangle$  
is a non-decreasing function of  the deposition  rate $w.$
 For small values of $w$, the system is likely to get arrested in some absorbing configuration  with 
 number of particles $N \le {\mu L \over \mu +1}$. Again, for $w \simeq 1$, the  density  fluctuates  about a large average value  $\rho(w)$ (note that  $\rho(1)=1$) and  the system remains active.  
 Thus, we expect  a transition from  the active phase to the absorbing state 
when $w$ is  decreased  below a  critical value  $w_c,$ such that  $\rho(w_c)\le  {\mu \over \mu +1}.$ 

\begin{figure}[t]
\includegraphics[width=8cm]{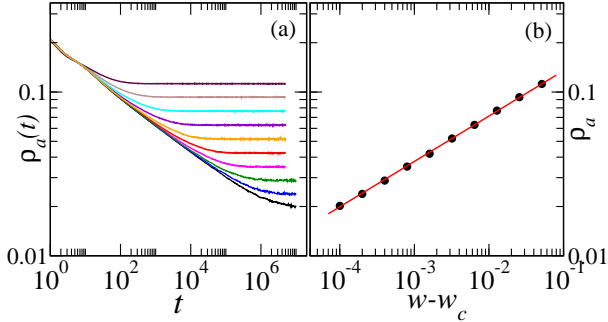}\vspace*{-0.3 cm}
\caption{The critical exponent $\beta$: (a) Plot of $\rho_a(t)$ vs. $t$ for different values of $w-w_c=0.0001,0.0002,..,0.0512$ for a system of size $L=2^{15}.$ (b) Corresponding saturation 
values are plotted against $w-w_c$ in log-scale.  The  best  fit straight line corresponds to 
the slope  $\beta = 0.278(1).$ }\vspace*{-0.3 cm}
\label{fig:beta}
\end{figure}

Let us  discuss  the $\mu=1$ case  in  details.  This exercise can be repeated straightforwardly  for any $\mu>1.$ The complete  dynamics for  RASEP with  $\mu=1$ is 
\bea
110  \rightarrow  101   & {\rm conserving}  \\
110  \mathop{\leftrightarrows}_{w}^{1-w}  111 & {\rm non-conserving}. 
\label{eq:110_101}
\eea
Any configuration devoid of $110$ is absorbing. 
In other words all the configurations  which have two or more consecutive $1$s are active.
Thus,    
\beq \rho_a= \langle s_{i-1} s_{i} (1-s_{i+1}) \rangle \eeq 
plays the role  of order parameter.  In the following we use Monte-Carlo simulations to  study  the  critical behaviour of this APT.

{\bf Critical point and decay exponent $\alpha$:}
Starting from a homogeneously active configuration, the activity $\rho_a(t) \sim t^{-\alpha}$
at the critical point $w_c.$ The propagation of activity is shown in the left panel of Fig. \ref{fig:clus}.
In the active phase $w > w_c$,  the activity eventually  saturates to some finite value  
and in the absorbing phase $\rho_a(t)$ vanishes  in the long time limit. One can then estimate the critical point $w_c$  and the exponent $\alpha$ by plotting $\rho_a(t)$ versus $t$ for different values of $w$ and looking for a power law decay. These estimates  can be verified by  plotting $\rho_a(t)t^\alpha$ against $t$ for different values of $w;$ the curve which remains horizontal in the long time limit corresponds to the critical point $w_c.$ This procedure, illustrated in  Fig. \ref{fig:wc}, leads to the estimates  
\bea
w_c=0.56757(1)~~ {\rm and }~~\alpha = 0.15984(7). \n
\eea
The decay exponent $\alpha$ is in very good agreement with the corresponding DP  value 
(see Table \ref{table}).  The density at the critical point is  
\begin{equation}
 \rho(w_c) = 0.4915(5), \n
\end{equation}
which is slightly lower than the  critical density $\rho_c= 1/2$  for  
conserved RASEP (i.e. when  dynamics (\ref{eq:110_101}) is  absent).    
The detailed  variation of  $\rho$ with $w$ will be discussed  later in this article. 

\begin{figure}
\vspace*{-0. cm}
\centering \includegraphics[width=6.5 cm,angle=0]{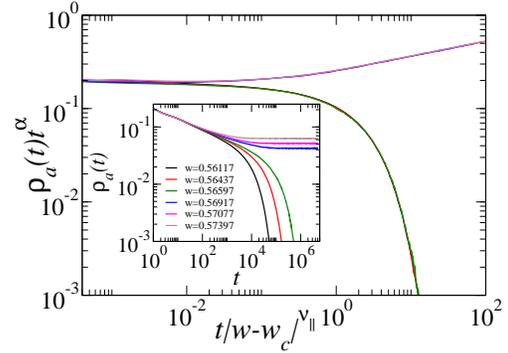}\vspace*{-0.3 cm}
\caption{The data collapse according to Eq. \eqref{eq:rhot_eps} for $|w-w_c| = 0.0016,0.0032,0.0064$
using  $\alpha=0.1598$ and $\nu_\shortparallel=1.74.$ The inset shows the  unscaled data.
The system size is $L=2^{15}.$}\label{fig:nu_par}\vspace*{-0.3 cm}
\end{figure}

{\bf The order parameter exponent $\beta$:}  
In the super critical regime $w> w_c,$ the activity saturates to some finite value $\rho_a.$
This saturation value shows a power law behaviour near the critical point,
\bea
\rho_a \sim (w-w_c)^\beta
\eea
where $\beta$ is the order parameter exponent.  Figure \ref{fig:beta}(a)
shows plot of $\rho_a(t)$ versus $t$ for various values of $w-w_c$ in the active phase. Corresponding saturation values
are plotted  against $w-w_c$ in Fig. \ref{fig:beta}(b). The slope of the resulting straight line gives an estimate 
\bea
\beta = 0.278(1), \n
\eea
consistent with $\beta_{DP}.$

{\bf Off-critical simulations and $\nu_\shortparallel$ :}
For $w \ne w_c$, the activity $\rho_a(t)$ obeys a scaling form 
\bea 
 \rho_a(t,w) = t ^{-\alpha} {\cal F}\left( t/\xi_\shortparallel  \right). 
\label{eq:rhot_eps}
\eea
where $\xi_\shortparallel$ is the characteristic time scale of the system. 
Near the critical point, both in the active ($w>w_c$) and absorbing ($w<w_c$) phase 
this scale is expected to diverge as
\bea
\xi_\shortparallel \sim |w-w_c|^{-\nu_\shortparallel}.
\eea
The exponent $\nu_\shortparallel$ can, thus, be estimated by plotting 
$\rho_a(t)t^{\alpha}$ versus $t |w-w_c|^{\nu_\shortparallel}$ for different values of $w$
and searching for a data collapse. This is  shown in  Fig. \ref{fig:nu_par}; the upper and lower branches correspond to $w>w_c$ and $w<w_c$ respectively. This gives us an estimate 
\bea
\nu_\shortparallel = 1.74(1), \n
\eea
which is also in good agreement with  the corresponding 
DP  value $1.733847(6).$

%

\begin{figure}
\centering \includegraphics[width=6.5 cm]{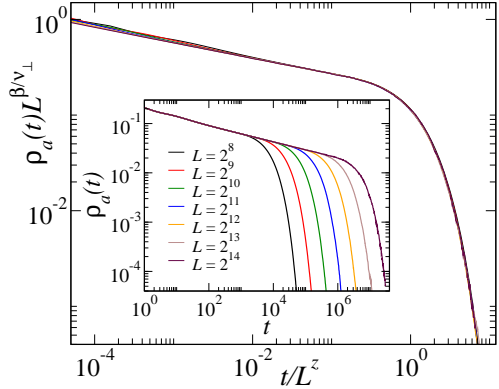}\vspace*{-0.2 cm}
\caption{Finite size scaling :$\rho_a(t)L^{\beta/\nu_\perp}$  as a function of the scaled variable $t/L^z$ for systems of sizes $L=2^8, 2^9,\dots  2^{14}$ are  collapsed using $\beta/\nu_\perp = 0.254(4)$ and  $z=1.58(1)$. The  unscaled data are shown  in the inset.}
\label{fig:z}\vspace*{-0.3 cm}
\end{figure}

{\bf Finite size scaling :} For a finite sized system the decay of activity $\rho_a(t)$ at the critical point $w_c$ is expected to follow a scaling form 
\bea 
\rho_a(t) = L^{- \beta/\nu_\perp} {\cal G}(t/L^z),
\label{eq:rhot_L}
\eea 
where $z$ is the dynamic exponent. Thus, at the critical point  the plots  of $\rho_a(t)L^{\beta/\nu_\perp}$ against $t/L^z,$  for different values of $L,$ are expected to  collapse onto a single function. 
The best data collapse, as shown in Fig. \ref{fig:z},  is obtained by using  
\bea
\beta/\nu_\perp = 0.254(4); \quad z=1.58(1). \n
\eea
 The inset there shows the  unscaled data  for  $L=2^8$-$2^{14}$. 
Both these estimates  are consistent with the DP exponents. It should be noted that this value of 
$\beta/\nu_\perp$ is strikingly different from that of the ordinary conserved RASEP, where $\beta/\nu_\perp =1$ 
was known exactly\cite{rasep}.

 The  estimated exponents  of non-conserved RASEP are summarized in  Table \ref{table}  
along with  the corresponding exponents  of  DP for comparison. All the exponents are in excellent 
agreement with those of DP, indicating that  RASEP  indeed flows to the DP universality class 
when the density conservation is broken.

In ordinary RASEP,   particle density was varied to observe an  absorbing transition 
at $\rho_c=1/2.$ It is therefore quite appropriate to ask whether   the average 
density $\rho$, which  now depends on $w$, can  also  be  treated as control parameter in the non-conserved version. To answer this question, let us first look at the behaviour of  $\rho$ near the critical point.


\section{The density $\rho(w)$} 
The average particle density $\rho$  is  a well defined 
function of  the deposition rate $w$ in the   supercritical  phase $w > w_c.$    Near the critical point, however,  like   all other observables the  density $\rho(w)$ is  expected  to have 
a non-analytic form,
\bea
\rho(w)=\rho(w_c)+ (w-w_c)^\kappa + {\rm higher ~ order ~ terms}, 
\eea 
where $\rho(w_c)$ is the  value to which the density saturates at the  critical point  and the exponent $\kappa$ is related to the critical exponents, possibly in a non-trivial way.

\begin{figure}
 \centering
\includegraphics[width=8.7 cm]{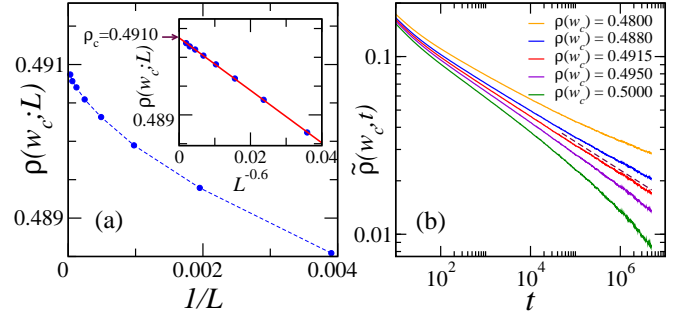}\vspace*{-0.3 cm}
 \caption{Determination of $\rho(w_c)$ :(a) Plot of $\rho(w_c;L)$ versus $1/L$ for $L=2^8-2^{15}.$ The inset shows plot of same date versus $L^{-0.6};$ linear fit yields $\rho(w_c)=0.4910.$ 
 (b) $\tilde\rho(t)$ is plotted against $t$ in log-scale for various trial values of $\rho(w_c);$
the best estimate $\rho(w_c)=0.4915(5)$ results in a straight line.}
 \label{fig:rhoc_L}\vspace*{-0.3 cm}
\end{figure}

To  estimate $\kappa$ we need to  determine $\rho(w_c)$ accurately. 
This can be done by extrapolating $\rho(w_c;L),$ the saturation value of density for a finite system of size $L,$ to 
the thermodynamic limit. As illustrated in Fig. \ref{fig:rhoc_L}, 
$\rho(w_c; L)$  approaches  a finite value  in this limit   $1/L \to 0.$  Since 
$\rho(w_c; L)$  is not   linear  in $1/L,$ it is useful to plot $\rho(w_c;L)$ against $L^{-b}$ with a suitable exponent $b$ that  gives a straight line.  The inset of Fig. \ref{fig:rhoc_L}(b) shows that  
$b=0.6$  does this job; a  linear  fit now gives an estimate $\rho(w_c)= 0.4910$. 

A more accurate estimate of $\rho(w_c)$ can be obtained from the  decay of $\rho(w_c,t).$ 
Since, the  characteristic timescale diverges at the critical point,  
in the  long time limit we expect   the  reduced density $\tilde \rho(w_c,t) \equiv (\rho(w_c,t)-\rho(w_c))$ to
decay  algebraically  $$\tilde \rho(w_c,t) \sim t^{-a}.$$

Then  $\rho(w_c)$ can be determined 
by plotting $\tilde \rho(w_c,t) $ versus $t$ in double logarithmic scale and adjusting the value of $\rho(w_c)$ in a way that the resulting curve is a straight line. This is shown in Fig. \ref{fig:rhoc_L}(a), 
and we have an estimate  $a= 0.1596(1)$ and $\rho(w_c) = 0.4915(5).$  
We  proceed further with  this  estimate  of $\rho(w_c).$

\begin{figure}
 \centering
 \includegraphics[width=8.7 cm]{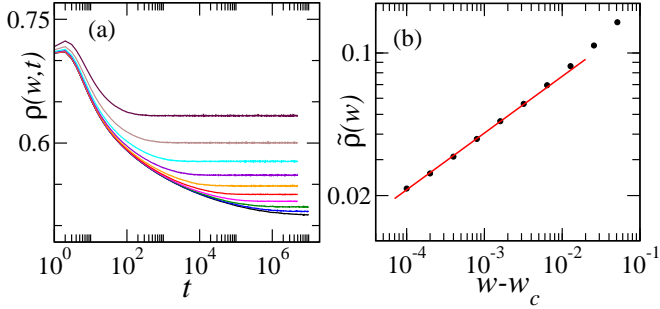}\vspace*{-0.3 cm}
 \caption{Determination of $\kappa$ : (a) The density $\rho(w,t)$ versus $t$ 
saturates to $\rho(w)$ for different $w-w_c=0.0001,0.0002,..,0.0512.$  (b) Corresponding  reduced density $\tilde \rho(w) =\rho(w)-\rho(w_c)$  vanishes as  $(w-w_c)^\kappa$ 
with $\kappa = 0.278(3).$ Here $L$ is $2^{15}.$}
 \label{fig:rho_beta}\vspace*{-0.3 cm}
\end{figure}

Determination of the exponent $\kappa$ is   described in Fig. \ref{fig:rho_beta}, where  
saturation  of $\rho(w,t)$ for different values of $w$  is shown in the left panel.  
Corresponding $\tilde\rho(w)$ obtained from this graph is plotted against $w-w_c$ in logarithmic  scale in Fig. \ref{fig:rho_beta}(b).  
The slope of this line over the first two decades corresponds to 
\bea
\kappa = 0.278(3). \n
\eea
Surprisingly, the exponents $a$  and $\kappa$, within the error bars,  are  identical to the DP exponents $\alpha_{DP}$ and $\beta_{DP}$  respectively.  This leads us to surmise that  $\tilde\rho(w)$ also undergoes an absorbing phase transition at  $w=w_c$ with  a decay exponent $\alpha=a$ and order parameter exponent $\beta=\kappa.$  We investigate  the other critical exponents   to confirm  that this transition, 
where  the reduced density $\tilde\rho(w)$ is  the order parameter,  also shows  DP behavior.  

\begin{figure}
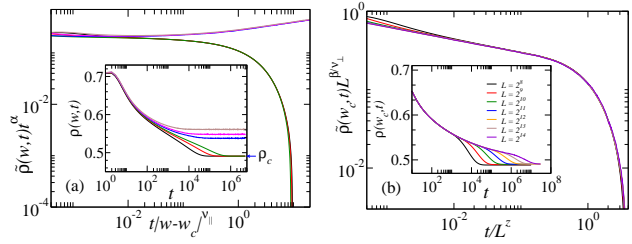

 
 \includegraphics[width=4 cm]{./rho_nupar.eps}$~$
 \includegraphics[width=4 cm]{./rho_z2.eps}
\vspace*{-.2 cm}
 \caption{
(a) $\tilde \rho(w,t)t^\alpha$ for different values of $w$ collapse  when plotted against $t|w-w_c|^{\nu_\shortparallel}$ yielding an estimate $\nu_\shortparallel = 1.73(1).$ 
(b) $\tilde \rho(w_c,t) L^{\beta/\nu_\perp}$ versus $t/L^z$ for different$L$ are collapsed to the 
finite size scaling function  for $z=1.58(1)$ and $\beta/\nu_\perp=0.256(4).$ The insets  of (a) 
and (b) show the unscaled $\rho(t)$ versus $t$ data.}
 \label{fig:rho_nupar}\vspace*{-0.3 cm}
\end{figure}

First, the off critical scaling.
In the inset of  Fig. \ref{fig:rho_nupar}(a) we have  plotted $\rho(w,t)$ versus  $t$ for different 
values of $w.$ Unlike  $w>w_c$ regime where  the saturation density $\rho(w)$ varies with $w,$ in  the 
sub-critical phase  $\rho(w<w_c)=\rho(w_c),$ ensuring that the order parameter $\tilde\rho(w)$
vanishes in this phase. The main figure Fig. \ref{fig:rho_nupar}(a) shows
the plot of $\tilde\rho(w,t)t^\alpha$ versus $t|w-w_c|^{\nu_\shortparallel}$ for different values of $w$, 
which could be collapsed using 
\bea
\nu_\shortparallel = 1.73(1). \n
\eea
This value of $\nu_\shortparallel$, again,  is compatible with DP. 

At the critical point, $\tilde\rho(w,t)L^{\beta/\nu_\perp}$  versus $t/L^z$ plot (see Fig \ref{fig:rho_nupar}(b)) for different values of $L$  collapse  following a  scaling form  similar to Eq. \eqref{eq:rhot_L}  
for
\bea
\beta/\nu_\perp=0.256(4); \quad z = 1.58(1). \n
\eea
All these exponents are listed in  Table \ref{table}. 
The fact that  these exponents  
are  consistent with   those  of $\rho_a$ as well as DP,  leads us to
conclude  that in absence  of particle conservation the reduced density $\tilde\rho(w)$ can  
be regarded as an alternative order parameter.  
  
\section{Density $\rho$ as a control parameter} 
 To have  a direct correspondence  with  the  APT in conserved RASEP, where  
 density was  tuned,  we   study the  transition   of the non-conserved model 
 considering  the average density $\rho$  as the tuning parameter. 
Evidently,  the critical point $w=w_c$    translates to 
$\rho_c \equiv \rho(w_c)$ in the density axis. 
In the following we argue that  the  critical exponents of this $\rho_a$ versus $\rho$ transition 
are  different from DP,  but they  can be expressed  in terms of DP exponents in 
a trivial way.

Along the density axis, the distance from the critical point  is $\Delta= \rho-\rho_c;$
since $\rho-\rho_c \sim (w-w_c)^\beta,$  we have   $(w-w_c) \sim  \Delta^{1/\beta}.$  
Therefore, critical exponents   corresponding to those   observables  which 
vary algebraically with $w-w_c$ near the critical point (e.g. $\rho_a \sim \Delta^{\beta'}$, $\xi_\shortparallel \sim \Delta^{-\nu_\shortparallel'}$  and  $\xi_\perp \sim \Delta^{-\nu_\perp'}$) are now rescaled by a  factor 
$1/\beta_{DP},$ 
\begin{equation}
 \beta'=  1; \quad \nu_\shortparallel' = \nu_\shortparallel/\beta; \quad \nu_\perp'= \nu_\perp/\beta.
\label{eq:exact}
\end{equation}
Obviously, the decay exponent $\alpha'=\beta'/ \nu_\shortparallel'$ and the dynamic exponent  
$z'=   \nu_\shortparallel'/\nu_\perp'$ do not change.

\begin{figure}
 \centering
 \includegraphics[width=8. cm]{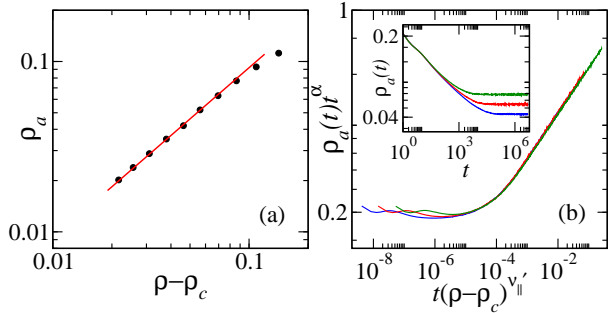}\vspace*{-0.3 cm}
 \caption{ Exponents of  $\rho_a$ vs. $\rho$ transition: 
(a) An excellent match of $\rho_a$ vs. $\rho-\rho_c$ data (symbols)  
with  a line  of unit slope  confirms  $\beta'=1.$ (b) Scaling collapse of $\rho_a(t)$  
following Eq. (\ref{eq:rhot_eps}), using $\nu_\shortparallel' 
= (\nu_\shortparallel/\beta)_{DP}.$  Inset of (b) shows unscaled $\rho_a(t).$ }
 \label{fig:rhoa_rho}\vspace*{-0.3 cm}
\end{figure}

The rescaling   of exponents Eq. \eqref{eq:exact} can be confirmed  by  re-plotting the 
data of Fig. \ref{fig:beta}(b) and Fig. \ref{fig:nu_par}   using $\rho(w) -\rho(w_c)$ instead of $w-w_c,$ which is shown 
in Fig.  \ref{fig:rhoa_rho}.  The left panel  of the figure  shows  $\rho_a$  as a function of 
$\rho -\rho_c$  resulting in the   order parameter  exponent $\beta'=1.$   Data collapse 
according to scaling form Eq. (\ref{eq:rhot_eps})  is shown in the right panel for the supercritical  regime 
$\rho>\rho(w_c)$  with  an unusually large $\nu_\shortparallel' = (\nu_\shortparallel/\beta)_{DP}=6.269.$   Note that   in this  scenario, sub-critical 
branch  does not exist as  the density $\rho$ does not vary  for $w<w_c.$

It is worth mentioning that although the critical exponents  of this transition, when  viewed  
by  tuning $\rho,$  are  widely different from  those of DP, the  basic underlying  
critical behaviour is DP.  It is only a coincidence that,  like the  conserved RASEP 
the order parameter $\rho_a$ here varies  linearly with $\rho-\rho_c$ near the critical 
point.

\section{Restricted Symmetric Exclusion Process}
Restricted Symmetric Exclusion Process (RSEP) is a variation of RASEP where a particle is allowed 
to move in forward (backward) direction  when it is preceded (followed) by at least $\mu$ particles. 
In particular, the dynamics for $\mu=1,$ which is identical to that of $1$-dimensional conserved lattice gas (CLG), is
\bea
110 \to 101 \leftarrow 011.\label{eq:110_011}
\eea
It was shown \cite{rasep, Oliveira} that this system also undergoes an APT  with integer exponents \eqref{eq:rasepexp} and belongs to the same universality class as RASEP. This non-DP behaviour seen  in one dimensional CLG is very different from  other  related models  with an additional conserved field,  
namely  conserved threshold transfer  process (CTTP) and conserved Manna  model though 
all three models    in two and higher dimensions   show   identical  critical  behaviour.  
In 1D, where CLG is exactly solvable, CTTP and  Manna  model  shows a nontrivial absorbing 
phase transition belonging to  DP.  It is natural to ask,  what 
makes CLG different from  others ? 

One possible  cause, like in RASEP, is  the   presence of  the conserved  density field.  
To confirm this  we   add  particle deposition and evaporation 
\bea
110 \mathop{\leftrightarrows}_{w}^{1-w} 111 \mathop{\leftrightarrows}_{1-w}^{w} 011,
\eea 
to the dynamics \eqref{eq:110_011} of CLG.
Clearly, this  dynamics preserves the absorbing configurations of the system, assuring that 
there is still  an absorbing state transition.  Detailed  analysis of the 
model  reveals that the transition occurring at $w_c=0.5635(1)$ is  identical to the  
that observed in  RASEP with $\mu=1,$  except that the  activity now propagates 
symmetrically  in space, looking  like a familiar  DP cluster (see Fig. \ref{fig:clus}). The critical exponents characterizing this transition also belong to the DP class; we skip the details of this study here.

\begin{center}
\begin{table}[t]\footnotesize
\begin{tabular}{|c|ccccc|}
\hline 
 APT &  $\alpha$ & $\beta$ & $\nu_\shortparallel$ \hspace*{-.2 cm}& $z$\hspace*{-.2 cm} & $\beta / \nu_\perp$ \hspace*{-.2 cm}\\
\hline
  DP \cite{dp} & 0.1594\hspace*{-.2 cm} & 0.2764\hspace*{-.2 cm} & 1.733\hspace*{-.2 cm} & 1.5802\hspace*{-.2 cm} & 0.2520 \hspace*{-.2 cm}\\
 $\rho_a$ vs. $w$  & 0.15984(7)\hspace*{-.2 cm} & 0.278(1)\hspace*{-.2 cm} & 1.74(1)\hspace*{-.2 cm} & 1.58(1)\hspace*{-.2 cm} & 0.254(4) \hspace*{-.2 cm}\\
 $\tilde \rho $ vs. $w$ &  0.1596(1)\hspace*{-.2 cm} & 0.278(3)\hspace*{-.2 cm} & 1.73(1)\hspace*{-.2 cm} & 1.58(1)\hspace*{-.2 cm} & 0.256(4) \hspace*{-.2 cm}\\
 $\rho_a$ vs. $\rho$ &-\hspace*{-.2 cm} & 1 \hspace*{-.2 cm}& 6.269\hspace*{-.2 cm} & -\hspace*{-.2 cm} & - \hspace*{-.2 cm}\\
\hline
\end{tabular}
\caption{Critical exponents of non-conserved RASEP.}\label{table}
\end{table}
\end{center}

\section{Summary}
Restricted exclusion processes are known to undergo absorbing phase transitions characterized by integer exponents even though the order parameter is a fluctuating scalar with no special symmetry except the presence of the additional conserved density field. This could have been blamed for the non-DP behaviour
but for the  examples of Manna model, CTTP etc. which show DP behaviour even in presence of a conserved field. In this work we show that when density conservation is violated,  by  allowing deposition
and evaporation of particles, the system  shows an APT at a finite deposition rate $w_c$.  The resulting 
critical behaviour  becomes  DP, confirming that the conservation of density was indeed   responsible for 
the non-DP behaviour in RASEP.

The density  $\rho(w)$ is non-analytic at the critical point and  we  find that $\rho-\rho(w_c)$ 
also  undergoes a transition  at $w=w_c$ with critical exponents identical (within error limits) 
to those  of DP.

The original transition in conserved RASEP was observed by tuning density. In the non-conserved 
version,  the activity  $\rho_a$ also show a transition  w.r.t.  $\rho$  at $\rho_c= \rho(w_c)$ 
with exponents  $\beta', ~ \nu_\shortparallel', ~ \nu_\perp'$ much  different   from  DP. We argue
that they are  in fact scaled by a factor $1/\beta_{DP}.$

The one dimensional CLG, which is  the symmetric  version of a 
special case  $\mu=1$ of RASEP,  also flows to DP when density conservation 
is broken.  In  a recent work, Fiore et. al. \cite{fiore} have claimed that  a modified 
CLG,  which includes   next nearest neighbour interaction and 
simultaneous jump of two  consecutive particles in opposite 
direction but does not violate  particle conservation, 
shows  non-trivial  critical behaviour. It still 
remains to see whether  it is possible to  modify  the dynamics  
of RASEP in a different way,  keeping density conservation, 
to obtain the DP critical behaviour.


\begin{thebibliography}{99}
\bibitem{dp}  M. Henkel, H. Hinrichsen, and S. L\"ubeck, {\it Non-Equilibrium phase transitions}, vol. 1, Springer, Berlin, 2008; H. Hinrichsen, Adv. Phys.\textbf{49}, 815 (2000).
\bibitem{gras} H. K. Janssen, Z. Phys. B \textbf{42}, 151 (1981); 
P. Grassberger, Z. Phys. B \textbf{47}, 365 (1982).
\bibitem{react-diff} J. Marro and  R. Dickman,
 {\it Nonequilibrium Phase transition in lattice models},  Cambridge University Press, Cambridge, 1999.
\bibitem{depinning} F. D. A. A. Reis, Braz. J. Phys. \textbf{33}, 501 (2003).
\bibitem{spread}  P. Grassberger, J. Stat. Phys. \textbf{79}, 13 (1995).
\bibitem{synchro} M. A. M\"unoz and R. Pastor-Satorras, Phys. Rev. Lett. \textbf{90}, 204101 (2003).
\bibitem{pk} P. K. Mohanty and D. Dhar, Phys. Rev. Lett. \textbf{89}, 104303 (2002).
\bibitem{ca} G. \'Odor and A. Szolnoki, Phys. Rev. E \textbf{  53}, 2231 (1996).
\bibitem{DPexp} K. A. Takeuchi, M. Kuroda, H. Chat\'e, and M. Sano, 
Phys. Rev. E \textbf{80}, 051116 (2009).
\bibitem{DPjensen} I. Jensen, J. Phys. A {\bf 32}, 5233 (1999).
\bibitem{cdp} J. W. Essam, J. Phys. A \textbf{22}, 4927 (1989).
\bibitem{BARW} H. Takayasu and A. Yu. Tretyakov, Phys. Rev. Lett., {\bf 68}, 3060 (1992);
 I. Jensen, J. Phys. A \textbf{26}, 3921 (1993).
\bibitem{Oliveira} M. J. de Oliveira, Phys. Rev. E \textbf{71}, 016112 (2005); 
E. F. da Silva, and M. J. de Oliveira, Jnl. Phys.   A : Math. Theor. {\bf 41}, 385004 (2008).
\bibitem{rasep} U. Basu and P. K. Mohanty, Phys. Rev. E \textbf{79}, 041143 (2009).
\bibitem{cons-h}  M. A. M\"unoz, R. Dickman, A. Vespignani and S. Zapperi, 
Phys. Rev. E \textbf{59}, 6175(1999);   R. Dickman, M. A. M\"unoz, A. Vespignani and S. Zaperi, 
Brazilian Jnl. Phys. \textbf{30}, 27 (2000).
\bibitem{clg} M. Rossi, R. Pastor-Satorras, and A. Vespignani, \prl {\bf 85}, 1803 (2000).
\bibitem{nomanna} M. Basu, U. Basu, S. Bondyopadhyay, P. K. Mohanty, and H. Hinrichsen, Phys. Rev. Lett. (2012).
\bibitem{cclf} U. Basu, M. Basu, and P. K. Mohanty, submitted to Phys. Rev. E.
\bibitem{Lee} S-G. Lee and S. B. Lee, Phys. Rev. E \textbf{77}, 021113 (2008).
\bibitem{fiore} C. E. Fiore and M. J. Oliveira, Braz. J. Phys. {\bf 36}, 218 (2006).
\end{thebibliography}
\end{document}